\documentclass[10pt,a4paper]{article}
\usepackage[textwidth=153mm,textheight=250mm,centering]{geometry}
\usepackage{graphicx}
\usepackage{xcolor}
\usepackage{booktabs}
\usepackage{array}
\usepackage{longtable}
\usepackage{float}
\usepackage{amsmath}
\usepackage{textcomp}
\usepackage[font=small,labelfont=bf,labelsep=period]{caption}
\usepackage[colorlinks=true,allcolors=blue]{hyperref}

\newcommand{\uV}{\,\textmu V}
\raggedright
\let\oldtabular\tabular
\renewcommand{\tabular}{\footnotesize\oldtabular}

\newcommand{\articletype}[1]{}
\renewcommand{\title}[1]{{\LARGE\bfseries\raggedright #1\par}\vspace{4mm}}
\renewcommand{\author}[1]{{\normalsize\raggedright #1\par}\vspace{2mm}}
\newcommand{\affil}[1]{{\footnotesize\raggedright #1\par}}
\newcommand{\email}[1]{\vspace{2mm}{\footnotesize\raggedright\textbf{E-mail:} #1\par}\vspace{2mm}}
\newcommand{\keywords}[1]{{\footnotesize\raggedright\textbf{Keywords:} #1\par}}
\newcommand{\orcid}[1]{\href{https://orcid.org/#1}{\includegraphics[width=8pt]{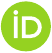}}}
\renewenvironment{abstract}{\vspace{4mm}{\color{gray}\hrule}\vspace{2mm}%
  \noindent{\bfseries Abstract}\par\vspace{1mm}\noindent\ignorespaces}%
  {\par\vspace{2mm}{\color{gray}\hrule}\vspace{4mm}}
\newcommand{\ack}[1]{\section*{Acknowledgments}#1}
\newcommand{\funding}[1]{\section*{Funding}#1}
\newcommand{\data}[1]{\section*{Data availability}#1}
\newcommand{\roles}[1]{\section*{Author contributions}#1}
\newcommand{\suppdata}[1]{\section*{Supplementary data}#1}

\begin{document}

\articletype{Paper}

\title{How large an effect would we have seen? Positive controls, detection floors and dataset pitfalls in scalp-EEG speech decoding}

\author{Blessing Isoyiza Adeika$^{1,*}$\orcid{0009-0007-5600-8051} and Onyema Osuagwu$^1$}

\affil{$^1$Morgan State University, Baltimore, Maryland, United States of America}

\affil{$^*$Author to whom any correspondence should be addressed.}

\email{blade3@morgan.edu}

\keywords{imagined speech, electroencephalography, electromyographic artifact, positive control, detection floor, brain--computer interface}

\begin{abstract}
\textit{Objective.} Studies that decode imagined speech from scalp electroencephalography (EEG) commonly validate their pipelines with a positive control, usually overt speech, and report null results without stating what effect size the pipeline could have detected.
\textit{Approach.} Using ds003626 (``Thinking out loud''), KaraOne and Track~3 of the 2020 International Brain--Computer Interface Competition, we audited decoding of overt speech, articulator preparation and imagined speech with artifact-channel subsets, band-limited decoding and trial selection, under a protocol that enforces split integrity by assertion. We injected class-dependent effects of known amplitude into the real recordings to measure the detection floor of five decoder families, and we checked the ds003626 labels against the dataset's acquisition protocol.
\textit{Main results.} On ds003626, two lip electrodes decoded overt speech at 0.7430 and the eight external channels at 0.7068, against 0.4231 for the 128-channel scalp montage, and no decodable signal survived in the 8--30\,Hz band. On the other two corpora, within-subject accuracy rose monotonically with frequency into the range dominated by muscle activity. Band-power logistic regression detected an injected 8--30\,Hz desynchronization of 0.54\uV\ under a criterion fixed in advance, whereas a compact convolutional network (EEGNet) detected nothing within subject up to 4.31\uV. In ds003626, the condition codes are the reverse of the intuitive reading, the class encoding changed between public releases, and condition is a deterministic function of time within each session.
\textit{Significance.} In these corpora the overt-speech positive control is dominated by myogenic signal and cannot validate a decoding pipeline. We recommend that speech-decoding studies report band-limited and artifact-channel accuracy, the direction of accuracy across frequency, and a measured detection floor.
\end{abstract}

\section{Introduction}

Speech neuroprostheses have advanced rapidly with intracranial recordings. Electrocorticography and intracortical arrays now decode attempted speech in people with paralysis at 62 to 78 words per minute and with vocabularies of up to 125,000 words \cite{anumanchipalli2019,moses2021,willett2023,metzger2023,card2024}, and these systems are being considered for clinical use in amyotrophic lateral sclerosis and the locked-in state \cite{chaudhary2016,vansteensel2016,vansteensel2023,silva2024}. Scalp electroencephalography (EEG) would make such a device non-invasive and widely available, which is why decoding imagined or covert speech from the scalp has attracted a sustained research effort \cite{herff2016,schultz2017,panachakel2021,lopezbernal2022}. Progress there has been much slower. Reviews describe small vocabularies, heterogeneous paradigms and results that are difficult to compare across studies \cite{panachakel2021,lopezbernal2022,jin2025,su2025}, and two recent re-evaluations that applied a common protocol across public corpora found within-category imagined-speech decoding at or near chance \cite{vandyck2026,khanday2026}.

Because imagined speech has no behavioral output, a decoding pipeline cannot be checked against what the participant actually intended. It is instead validated indirectly, most often by showing that it decodes overt speech or the contrast between speaking and rest. That inference is sound only if the signal the pipeline finds in the control condition is cortical. Articulation produces electromyographic (EMG) activity at the scalp that is far larger than the cortical signal of interest \cite{devos2010,porcaro2015}. Its spectrum is broad, overlaps the high-frequency range of neural activity \cite{goncharova2003,muthukumaraswamy2013}, and under neuromuscular blockade most scalp power above 20\,Hz disappears \cite{whitham2007}. Ocular micro-saccades can mimic induced gamma-band responses \cite{yuvalgreenberg2008}, and even silent verbal tasks can raise orofacial muscle activity \cite{mcguigan1970,nalborczyk2017}. A positive control can therefore succeed for reasons unrelated to cortical speech processing. For the intended users, people who cannot articulate at all, a decoder that relies on articulator activity would not work \cite{brumberg2010}.

A second gap concerns null results. Decoding studies usually report accuracy and, at best, whether it exceeds chance \cite{combrisson2015,ojala2010}, but rarely state how large an effect the pipeline could have found. Evaluation is itself fragile: information can leak between training and test data in several ways \cite{kaufman2012,kapoor2023,brookshire2024}, selection performed outside cross-validation inflates accuracy \cite{kriegeskorte2009,whelan2014}, and cross-validated estimates from small samples carry wide error bars \cite{varoquaux2018}. Simulation frameworks that place known sources into EEG are well established for method development \cite{krol2018,barzegaran2019}, yet they are seldom used to attach a sensitivity bound to a decoding null.

In this paper we address three questions. First, is the overt-speech positive control neural in the corpora most often used for imagined-speech decoding? Second, what effect sizes can commonly used decoders detect in these recordings, and how does this depend on the type of effect? Third, do the labels of a widely used corpus mean what they appear to mean? We answer the first with channel-subset, band-limited and trial-selection controls on three corpora, the second by injecting effects of known amplitude into the real recordings, and the third by checking ds003626 against its acquisition protocol. The same evaluation protocol underlies a companion paper that reports cross-subject decoding of imagined-speech content on these corpora \cite{companionA2}.

\section{Related work}

\subsection{Imagined-speech decoding from scalp EEG}

Early work used common spatial patterns to discriminate imagined /a/, imagined /u/ and a no-imagery control, reporting 68--78\% accuracy for three participants \cite{dasalla2009}. Open corpora followed. KaraOne records EEG, facial motion and audio while participants imagine and then articulate seven phonemes and four words chosen as phonological minimal pairs; its authors reported up to 90\% accuracy for separating imagined consonants from vowels and up to 95\% for separating the stimulus period from active imagination, using deep belief networks on engineered features \cite{zhao2015}. Later work on the same corpus reported a mean of 83.42\% over six binary phonological tasks \cite{saha2019}. A corpus of five imagined Spanish vowels and six command words is also openly available \cite{coretto2017}. The ds003626 corpus, ``Thinking out loud'', records inner, pronounced and visualized versions of four Spanish direction words from ten participants \cite{nieto2022,nieto2022data}; on its four-class inner-speech task, EEGNet reached 29.7\% \cite{vandenberg2021} and a bidirectional recurrent network 36.1\% \cite{gasparini2022}. Track~3 of the 2020 International Brain--Computer Interface (BCI) Competition asked entrants to classify five imagined phrases, and the winning entry reached 82.6\% with the second at 75.5\% \cite{jeong2022}.

Two recent evaluations reached more sobering conclusions. Van Dyck applied seven decoders under one nested cross-validation protocol to seven public and two in-house corpora and found within-category imagined-speech decoding mostly at or near chance, with a best pairwise balanced accuracy of 62.7\% on the 2020 competition data \cite{vandyck2026}. Large EEG models pretrained on thousands of hours of recordings have been proposed as a route to generalization \cite{kostas2021,jiang2024}, but a benchmark on speech found that they offered no consistent advantage over a small convolutional network, and that every model tested was at chance on the competition data when the test participant was held out \cite{khanday2026}. Differences in evaluation protocol are a plausible source of this spread, which is why the protocol used here is described in detail (section~\ref{sec:protocol}).

\subsection{Myogenic and ocular contamination}

Muscle activity is the best documented contaminant of scalp EEG. Frontalis and temporalis contraction produces broadband scalp activity from below 20\,Hz to beyond 200\,Hz, detectable at the vertex even at low contraction levels \cite{goncharova2003}. Recordings made during neuromuscular paralysis show that most scalp power above 20\,Hz in the awake state is of muscular origin \cite{whitham2007}, and muscle activity can masquerade as neural effects down to the alpha band, with independent component analysis removing it adequately in some protocols but not others \cite{mcmenamin2010}. A review of high-frequency activity in MEG and EEG concluded that the spectra of muscle and neural activity overlap entirely and recommended specific controls \cite{muthukumaraswamy2013}. Miniature saccades generate spike potentials that appear as induced gamma-band responses \cite{yuvalgreenberg2008}.

Speech production magnifies the problem. Articulation artifacts in EEG recorded during overt naming and spoken language production exceed the brain signal and require dedicated separation methods \cite{devos2010,porcaro2015}, and EEG studies of language production have long had to work around them \cite{ganushchak2011}. Silent-speech interfaces show that the same articulator activity carries linguistic information when it is recorded directly \cite{schultz2017,kapur2018}. Whether inner speech itself engages the articulators is less settled: induced verbal rumination raises lip and forehead EMG \cite{nalborczyk2017,mcguigan1970}, but phonetic content could not be decoded from surface EMG during inner speech \cite{nalborczyk2020}. The neural signature expected from speech motor cortex is different in kind: a band-limited decrease of mu (8--13\,Hz) and beta (13--30\,Hz) power over sensorimotor regions \cite{pfurtscheller1999,jenson2014}. In intracranial recordings, high-frequency activity best decodes overt speech, whereas low-frequency power and cross-frequency coupling contribute to decoding imagined speech \cite{proix2022}; spectrotemporal features and word pairs of covert speech have also been decoded from high-gamma activity recorded directly from cortex \cite{martin2014,martin2016}.

\subsection{Reference choice and volume conduction}

The common average reference assumes dense, even scalp coverage and redistributes activity from every electrode into every other \cite{dien1998,nunez2006}. Alternatives such as the reference electrode standardization technique approximate a reference at infinity \cite{yao2001}. For connectivity analysis the consequence is well known: a shared reference and volume conduction create zero-lag coupling between electrodes that share no source, and measures that discard the zero-lag component, such as the imaginary part of coherency and the weighted phase-lag index, were introduced to address it \cite{nolte2004,stam2007,vinck2011,schoffelen2009,bastos2016}.

\subsection{Evaluation, leakage and sensitivity}

Leakage between training and test data is now recognized as a pervasive problem in machine-learning-based science \cite{kaufman2012,kapoor2023}. In EEG, splitting segments of the same recording across training and test inflates deep-learning accuracy \cite{brookshire2024}, and block designs allow classifiers to read slow temporal drift in place of the experimental variable \cite{li2021}. Reviews of deep learning for EEG report inconsistent evaluation and limited reproducibility \cite{roy2019}. On the statistical side, theoretical chance levels are misleading for small test sets, and binomial thresholds or permutation tests are required \cite{combrisson2015,ojala2010}. What is missing from most imagined-speech studies is a measured sensitivity: simulation toolboxes place sources of known strength into realistic EEG \cite{krol2018,barzegaran2019}, but their outputs are rarely used to state the detection floor of a reported null.

\section{Materials and methods}

\subsection{Corpora and trial structure}\label{sec:data}

Table~\ref{tab:corpora} summarizes the three corpora. All are openly available and were collected under the ethical approvals documented in their original publications \cite{nieto2022,zhao2015,jeong2022}.

\begin{table}[t]
\caption{The three corpora and the trial phases analyzed. Trial counts per class are per participant. The KaraOne recordings contain 68 channels typed as EEG, six of which are not cortical.}
\label{tab:corpora}
\centering
\setlength{\tabcolsep}{4pt}
\begin{tabular}{@{}p{22mm}p{6mm}p{35mm}p{28mm}p{47mm}@{}}
\toprule
Corpus & $N$ & Conditions and classes & Channels & Trial phases and analysis window \\
\midrule
2020 BCI Competition, Track 3 & 15 & imagined; 5 phrases (\textit{Hello, Help me, Stop, Thank you, Yes}); 80 per class & 64 EEG; 256\,Hz & epochs of 3.1\,s as distributed by the organizers, analyzed whole (0.0--3.1\,s) \\[2pt]
ds003626 & 10 & pronounced, inner, visualized; 4 Spanish words (\textit{arriba, abajo, derecha, izquierda}) cued by arrows; about 28 (pronounced) or 56 (inner) per class & 128 EEG and 8 external (EXG); 256\,Hz derivatives & epochs $-$0.5--4.0\,s; visual cue 0.0--0.5\,s; action interval 1.0--3.5\,s, analyzed \\[2pt]
KaraOne & 14 & imagined, and the interval before imagery; 7 phonemes and 4 words; about 12 per class for the imagined words & 62 scalp, plus 2 mastoid, 2 ocular, 1 cardiac and 1 EMG; 1000\,Hz & rest, stimulus with articulator positioning, imagined speech (median 4.985\,s here), speaking aloud; imagined 0.0--5.0\,s, pre-imagery 0.0--2.0\,s \\
\bottomrule
\end{tabular}
\end{table}

In ds003626 each trial begins with a visual arrow that cues one of the four words, followed by an interval in which the participant pronounces, imagines or visualizes it \cite{nieto2022}. We analyzed the action interval, 1.0--3.5\,s after cue onset, and used the cue interval, 0.0--0.5\,s, as a visual control. Each KaraOne trial passes through four states: rest; a stimulus state, in which the prompt is shown and played and which ends with a 2\,s period in which participants move their articulators into position; imagined speech; and speaking aloud \cite{zhao2015}. Besides the imagery interval we analyzed the 2.0\,s interval that precedes it, which we call the pre-imagery interval. It contains perception of the prompt and preparation to articulate it, but neither imagery nor speech (section~\ref{sec:karaone}). KaraOne's spoken productions, which follow imagery, were not analyzed. The competition corpus is distributed as fixed epochs of 3.1\,s. In the BIDS release used here these epochs are concatenated back to back with no inter-trial signal, which matters for filtering (section~\ref{sec:preproc}).

\subsection{Processing and modelling pipeline}\label{sec:pipeline}

Every analysis in this paper passes through the same sequence of stages, summarized in table~\ref{tab:pipeline} and detailed in the subsections that follow. The stages are ordered so that nothing estimated from data is fitted before the split that defines training and test data, and the last column of the table records, for each stage, which data any estimated parameter is fitted on. A stage that fits nothing cannot leak information; a stage that fits on training rows only cannot leak it across the split; the only stage that uses data from a held-out participant without labels is per-participant re-centering, and runs that use it are flagged.

\begin{table}[t]
\caption{Processing and modelling stages, in the order applied. ``Fitted on'' states which data any estimated parameter is computed from.}
\label{tab:pipeline}
\centering
\setlength{\tabcolsep}{4pt}
\begin{tabular}{@{}p{4mm}>{\raggedright\arraybackslash}p{27mm}>{\raggedright\arraybackslash}p{68mm}>{\raggedright\arraybackslash}p{38mm}@{}}
\toprule
 & Stage & Operation and settings & Fitted on \\
\midrule
1 & Loading and label checks & read derivatives (ds003626) or raw recordings (KaraOne, competition); normalize class codes across dataset releases; verify condition codes against the acquisition code (section~\ref{sec:integrity}) & nothing fitted \\
2 & Trial identity & identifier of corpus, participant, session, trial index, condition and prompt for every trial & nothing fitted \\
3 & Filtering and reference & 60\,Hz notch, 1--45\,Hz band-pass, common average reference, resampling to 128\,Hz; 512\,Hz, 1--200\,Hz and 256\,Hz, 1--100\,Hz variants for the high-frequency analyses & nothing fitted (fixed filters) \\
4 & Windowing & analysis window fixed per corpus before any decoding (table~\ref{tab:corpora}); band-limiting by per-trial spectral masking where required & nothing fitted \\
5 & Channel selection & subsets defined from montage geometry (ds003626: angle from the vertex, $\leq 46^{\circ}$ for the central cap, $\geq 80^{\circ}$ for the rim) or from channel type (KaraOne); KaraOne scalp channels re-referenced to their own mean & nothing fitted (fixed lists) \\
6 & Split generation & one module builds every split; validity assertion; split hash (section~\ref{sec:protocol}) & identifiers only \\
7 & Normalization & networks: z-score per trial and channel; feature models: standardization & within each trial; training rows \\
8 & Features & log band power per channel and band; spatial covariance with principal component reduction, shrinkage and re-centering; phase-lag, coupling and phase features & training rows; re-centering per participant from that participant's own unlabeled trials \\
9 & Model fitting & logistic regression ($C$ fixed); networks with early stopping on validation macro-F1 (section~\ref{sec:decoders}) & training rows; validation set drawn from training data \\
10 & Evaluation & held-out trials scored once; binomial threshold, Wilcoxon test, bootstrap interval, FDR correction, shuffle-label control (section~\ref{sec:stats}) & test rows, scoring only \\
11 & Verification and provenance & seven checks for unexpectedly high results; append-only run record with code version, data checksum, seed and split hashes & nothing fitted \\
\bottomrule
\end{tabular}
\end{table}

\subsection{Preprocessing}\label{sec:preproc}

For ds003626 we used the dataset authors' derivatives, which are already filtered and cleaned by independent component analysis \cite{nieto2022}, resampled to 128\,Hz. KaraOne and the competition data were built from the raw recordings with MNE-Python \cite{gramfort2013}: a 60\,Hz notch filter, a common average reference, a 1--45\,Hz band-pass filter and resampling to 128\,Hz. Analysis windows were fixed in advance and never chosen on accuracy. To examine frequencies above 45\,Hz we built a 512\,Hz, 1--200\,Hz version of the KaraOne pre-imagery data and a 256\,Hz, 1--100\,Hz version of the competition data. Band-limited decoding used brick-wall masking of the discrete Fourier transform within each epoch, applied identically to every band. For the competition corpus this per-trial approach is required: a 1\,Hz high-pass filter has an impulse response of roughly 3.3\,s, longer than one 3.1\,s trial, so filtering the concatenated epochs would mix each trial with its neighbors, which fall into different cross-validation folds.

\subsection{Evaluation protocol and split integrity}\label{sec:protocol}

\begin{figure}[t]
\centering
\includegraphics[width=\textwidth]{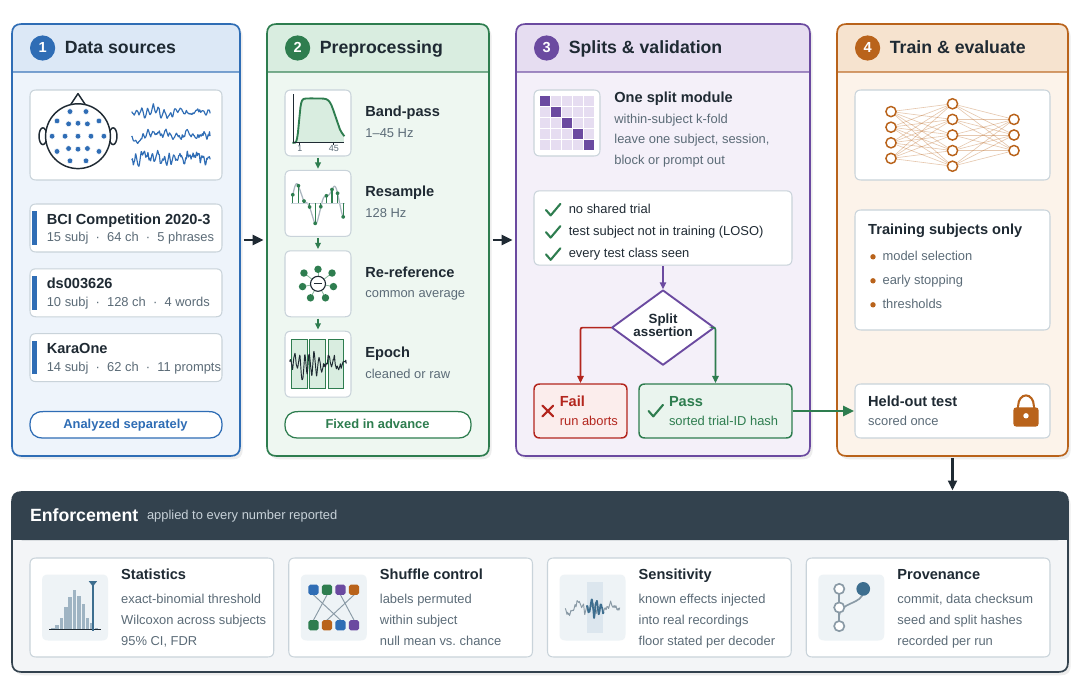}
\caption{The evaluation protocol. (1) The three corpora are analyzed separately. (2) Preprocessing is fixed in advance; the dataset authors' cleaned epochs are used where a corpus provides them, and epochs are otherwise built from the raw recordings. The analyses above 45\,Hz use 256 and 512\,Hz versions of the data (section~\ref{sec:preproc}). (3) Every trial carries an identifier built from its provenance (corpus, participant, session, trial index, condition and prompt), and one module generates every split. An assertion halts the run if a trial is shared between sets, if the test participant appears in training under leave-one-subject-out evaluation, if a test class is absent from training, or if the validation set is empty; a split that passes emits a hash of its sorted trial identifiers. (4) Model selection, early stopping and thresholds use training data only, and the held-out test set is scored once. Every reported value then carries the statistics, shuffle-label control, measured detection floor and provenance record listed in the lower panel. FDR, false discovery rate; CI, confidence interval.}
\label{fig:protocol}
\end{figure}

Figure~\ref{fig:protocol} summarizes the protocol. Every trial carries an identifier built from provenance alone: corpus, participant, session, trial index, condition and prompt. A single software module constructs every split, and before the first optimizer step an assertion stops the run if any identifier appears in more than one of the training, validation and test sets, if the test participant appears in training data under leave-one-subject-out evaluation, if a class present in the test set is absent from training, if the test set contains a single class, or if a model that performs selection has no validation data. A SHA-256 hash of the sorted identifiers is recorded for every fold, so that any change to the split machinery that alters a split is detectable. Early stopping, checkpoint selection and decision thresholds draw only on training data. We report within-subject stratified 5-fold cross-validation, leave-one-session-out and leave-one-block-out evaluation within subject, and leave-one-subject-out (LOSO) evaluation across participants.

Results that exceed the ranges expected for this problem were put through seven further checks before being reported: a shuffle-label control, a rebuild of the data cache with three fresh split seeds, replication on a second corpus, physiological plausibility of the learned spatial filters and frequency profile, decoders restricted to artifact channels, the highest band, the pre-cue baseline and the trial index, graceful degradation under channel and trial ablation, and the expected ordering of protocols and conditions. A result counted as verified only if all seven passed. Every run writes an append-only record containing the code version, a checksum of the input data, the random seed, the per-fold split hashes and the model configuration. The project archive holds 1,890 such records. The number of computational threads was fixed at four for all runs, because thread count alone changed accuracy by approximately 0.0018 through floating-point reduction order.

\subsection{Decoders}\label{sec:decoders}

A null result is informative only about the decoder that produced it, so we used decoders that differ in what they can represent: a linear model on band power, which reads amplitude changes directly; convolutional networks, which learn spatial and temporal filters from the raw signal; covariance-based classifiers on the Riemannian manifold, which capture cross-channel structure; and linear models on phase and coupling features, which capture timing relations that band power ignores.

The band-power decoder computes the logarithm of mean in-band power for each channel and band over the analysis window, standardizes the features with statistics fitted on training rows only, and applies L2-regularized multinomial logistic regression with the inverse regularization strength fixed at 0.1 \cite{pedregosa2011}. It performs no model selection and therefore needs no validation set. Features were computed either in single bands or in a six-band bank (1--4, 4--8, 8--13, 13--20, 20--30 and 30--45\,Hz).

EEGNet \cite{lawhern2018} was run with a temporal kernel of 64 samples and two filter configurations: 16 temporal filters with a depth multiplier of 4 ($F_1 = 16$, $D = 4$), and a smaller configuration with 8 temporal filters and a depth multiplier of 2 ($F_1 = 8$, $D = 2$). We also implemented ShallowConvNet \cite{schirrmeister2017} with seven temporal filters of 25 samples, a spatial convolution, and the square, mean-pool (75 samples, stride 15) and logarithm sequence that makes it an explicit band-power estimator. All networks were trained in PyTorch with the Adam optimizer (learning rate $10^{-3}$, weight decay $10^{-4}$), batches of 32, dropout of 0.3 and at most 100 epochs, on inputs z-scored per trial and channel. Twenty percent of the training data, drawn from training participants under LOSO, formed a validation set; validation macro-averaged F1 was evaluated every five epochs and training stopped after eight evaluations without improvement.

The Riemannian decoders estimate a spatial covariance matrix per trial, reduce it with a principal component projection fitted on training covariances, apply shrinkage toward a scaled identity \cite{ledoit2004}, re-center each participant's matrices at their Riemannian mean \cite{zanini2018}, and classify with the minimum distance to Riemannian mean, its geodesically filtered variant, or logistic regression in the tangent space \cite{barachant2010,barachant2012,congedo2017}, implemented with pyRiemann \cite{pyriemann}. The phase decoders apply the same logistic regression to three feature families computed per trial, band and region pair: the signed weighted phase-lag index \cite{vinck2011}, the complex phase--amplitude coupling vector with phase from theta and amplitude from beta \cite{tort2010,canolty2006}, and the phase of each region's analytic signal at a reference latency, encoded as its cosine and sine. The regions were a language set comprising left inferior frontal and left superior temporal cortex with their right homologues, and a central sensorimotor cap.

\subsection{Statistical analysis}\label{sec:stats}

For every result we report the number of test trials, chance level, the exact binomial significance threshold for a single participant at $\alpha = 0.05$ \cite{combrisson2015}, the number of participants at or above that threshold, a one-sided Wilcoxon signed-rank test of participant accuracies against chance \cite{wilcoxon1945}, and a bias-corrected and accelerated bootstrap 95\% confidence interval (CI) on the mean over 10,000 resamples \cite{efron1987}. A participant exactly at the threshold counts as clearing it, because the threshold is defined as the smallest accuracy whose one-sided $p$-value does not exceed $\alpha$. Families of related tests were corrected by controlling the false discovery rate (FDR) with the Benjamini--Hochberg procedure \cite{benjamini1995}.

Where a result is described as surviving its shuffle-label control, labels were permuted within each participant and the identical pipeline was run 200 times, or 50 times for the ds003626 network cells, where every draw retrains the network; permuting across participants would also destroy the structure that distinguishes participants and could inflate the null for reasons unrelated to the class labels. The observed value was recomputed by the same code path as the null draws. A result was rejected if the mean of the null distribution was significantly above chance or if the observed value did not exceed the 95th percentile of the null. Empirical $p$-values were computed as $(b + 1)/(m + 1)$ for $b$ null draws at or above the observed value out of $m$ \cite{phipson2010}.

\subsection{Artifact audit of the positive controls}\label{sec:audit}

The discriminating features of cortical speech-motor activity and of muscle activity are frequency and location: the former is a mu and beta power decrease over central sites \cite{pfurtscheller1999,jenson2014}, the latter is broadband from about 20\,Hz upward, increases with contraction and is largest at the periphery of the montage \cite{goncharova2003,muthukumaraswamy2013}. We therefore decoded each positive control from restricted sets of channels, bands and trials.

For ds003626 pronounced speech we compared the 128-channel scalp montage with a central sensorimotor cap of 40 channels, a peripheral and temporal rim of 52 channels, all eight external (EXG) channels, the two lip electrodes over orbicularis oris (EXG7--8), the ocular channels (EXG3--6, and their horizontal and vertical pairs separately), the earlobe reference pair (EXG1--2), and all EXG channels during the cue interval. We also decoded each of five bands separately (1--8, 8--13, 13--20, 20--30 and 30--45\,Hz), and repeated the decoding after aggressive amplitude and gradient rejection and on the cleanest and dirtiest halves of each participant's trials. These cells used EEGNet ($F_1 = 16$, $D = 4$) within subject with 5-fold cross-validation.

For phoneme decoding in the KaraOne pre-imagery interval, where a dedicated EMG channel exists, we compared all 68 stored channels, the 62 scalp channels re-referenced to their own mean, a 21-channel frontocentral speech-motor strip, a 22-channel peripheral rim matched in size, the non-scalp channels, the EMG channel alone and the ocular channels, and decoded five bands at 128\,Hz and three bands (8--30, 30--45 and 45--100\,Hz) at 512\,Hz. These cells used EEGNet ($F_1 = 16$, $D = 4$) and band-power logistic regression, within subject and under LOSO. For the competition corpus, which has no EMG channel, we decoded single bands up to 100\,Hz with band power.

\subsection{Injected effects of known size}\label{sec:inject}

To measure what each decoder could detect, we injected class-dependent effects into the real ds003626 inner-speech trials. The amplitude effect suppresses the participant's own ongoing 8--30\,Hz rhythm by a multiplicative depth instead of adding a synthetic oscillator, so the injected effect inherits the real phase statistics and spatial covariance of each recording. Per-participant rotation and tilt of the injected topography, a per-class positional offset, and amplitude and latency jitter across participants and trials ensure that the mapping from class to topography differs between people, as it would for a real effect. The effect is scaled by a factor $k$ times the mean 8--30\,Hz standard deviation over the central sensorimotor region, which is 5.20\uV\ in these recordings, times each participant's amplitude scale, whose mean is 1.036. One unit of $k$ therefore corresponds to 5.392\uV\ at the peak of a desynchronization sustained for 1.5\,s, and $k = 0.1$ is a 10\% reduction of the ongoing rhythm, or 0.54\uV. We swept $k$ from 0 to 0.8. A decoder was considered to detect an effect only if, under a criterion fixed before the sweep, its mean accuracy reached the mean per-participant threshold, at least eight of ten participants individually cleared their threshold, and the group Wilcoxon test gave $p < 0.05$; where several seeds were run, every seed had to meet the criterion.

We extended the injection to four effect types and all three corpora: the amplitude effect above, an evoked deflection, class-dependent phase locking, and class-dependent coupling between theta phase and beta amplitude. This grid of 240 cells was evaluated with band power and with a Riemannian tangent-space decoder, under both protocols, at five values of $k$. Because accuracy on real data can sit above chance before any injection, a grid cell counted as detected when participant accuracies exceeded those of the same cell at $k = 0$ under a one-sided paired Wilcoxon test with a mean gain above one percentage point, after Benjamini--Hochberg correction across all 240 cells.

Finally, two effects were injected specifically to validate the phase features: a class-dependent phase lag between two regions, obtained by replacing a fraction $k$ of one region's in-band component with a lagged copy of the other's, and a class-dependent coupling phase, with $k$ the depth to which theta phase modulates the beta envelope. Both preserve band power by construction. The phase-locking value \cite{lachaux1999} was computed alongside the lagged measures as a negative control, because a common average reference manufactures zero-lag phase agreement between electrodes that share no source \cite{nolte2004,bastos2016}.

\subsection{Integrity checks on the dataset labels}\label{sec:integrity}

For ds003626 we compared the condition codes read by our loader with the dataset authors' acquisition code and events decoder, the block structure of every session, the authors' EMG rejection flags, and the EXG power in each condition. We also compared the class encodings of public releases 1.0.0 and 2.1.2 of the dataset, and we tabulated the within-session position of every trial by condition. For KaraOne we compared the duration and position of every annotated segment, in all 14 participants, with the trial structure described by the dataset's authors \cite{zhao2015}.

\section{Results}

\subsection{Three ways ds003626 is misread}\label{sec:ds003626}

The condition codes of ds003626 are \texttt{0}~=~pronounced, \texttt{1}~=~inner and \texttt{2}~=~visualized. Because the corpus is named for inner speech, reading code \texttt{0} as inner speech is natural, and our own loader did so for the first months of this project. The correct mapping is established independently by the acquisition protocol's stimulation code, whose paradigm order contains one pronounced run followed by two inner and two visualized runs; by the observed 40/80/80 block structure in all 30 sessions; by the authors' events decoder and loading interface; by their EMG control, which excludes condition~0 by name and contains none of the 115 flagged trials in it; and by EXG power between 20 and 45\,Hz, which is elevated approximately ninefold in condition~0 in all ten participants. Under the corrected mapping, inner speech comprises 2,236 trials, not 1,128, and the within-subject accuracy of 0.342 that we had previously reported as inner speech is an overt-speech result.

The class encoding changed between public releases. In version 1.0.0 the four classes are coded 1 to 4, and in version 2.1.2 they are coded 0 to 3, with class identity and order preserved. A pipeline written for the earlier release that selects classes 1 to 4 silently drops one class and retains exactly 75\% of every condition when run on the current release, without raising an error. We normalize both encodings at every point where the events are read and guard the conversion with a regression test.

Condition is also collinear with time. Across all 30 sessions, pronounced trials occupy within-session positions 0 to 39, inner trials 40 to 119 and visualized trials 120 to 199. A single threshold on trial position separates inner from visualized trials with an accuracy of 1.0000 in every session, so condition is a deterministic function of time within session and cannot be separated from it by regression. Decoding inner versus visualized speech within subject reached 0.8604 before we recognized this. The only available control is geometric: comparing the last $w$ trials of one block with the first $w$ of the next keeps the groups adjacent in time while preserving the condition difference. Because a block boundary is also a run boundary, the control must itself cross a run break with condition held constant, and that calibrating contrast decoded at 0.5029 to 0.6333 on its own. Under a paired per-participant test, the inner-versus-visualized contrast survived correction at only one of four window widths, whereas the overt contrast survived at all four. This is the block-design confound described for EEG classification elsewhere \cite{li2021}, and it means that the condition axis of ds003626 cannot serve as a covert-speech control.

\subsection{A KaraOne interval annotated as speech contains none}\label{sec:karaone}

KaraOne's annotation file lists two speaking segments per trial in a single alternating sequence, one between rest and imagery and one after imagery. Only the second is speech. The first lasted a median of 2.00\,s in every participant and ended where imagery began, whereas the second varied in length from trial to trial, as a self-paced response does. The authors' protocol places the speaking state after imagery and, before imagery, a stimulus state that ends with a 2\,s period in which participants move their articulators into position \cite{zhao2015}. The first segment is therefore the pre-imagery interval: the prompt has been presented and the articulators are being positioned, but nothing is spoken. Our conversion of the recordings to the BIDS format took the first segment of each pair and exported it as spoken speech, and we analyzed it under that name until we checked it against the protocol.

The alternation also breaks in three participants. In one, 40 trials share an onset with another trial and carry two prompts. In two others the sequence is offset by one entry from some point in the recording onwards, so that 138 of the 1,781 segments exported for the other 13 participants (20 in one participant and 118 in the other) are the spoken response of the preceding trial, labelled with the current trial's prompt. We therefore took the pre-imagery interval directly as the gap between the end of rest and the start of imagery, which is unambiguous in all 14 participants, and kept a trial only if that gap was longer than the 2.0\,s analysis window, so that no window reaches into imagery; this removed 43 of 1,913 trials and left 1,186 phoneme trials. Run on the exported onsets for two participants, the same code reproduced the earlier export exactly, so the two versions differ only in where each window starts. Every KaraOne pre-imagery result in this paper and in the companion paper \cite{companionA2} uses the rebuilt interval.

\subsection{Within subject, the positive controls are dominated by myogenic signal}\label{sec:myogenic}

\begin{figure}[t]
\centering
\includegraphics[width=\textwidth]{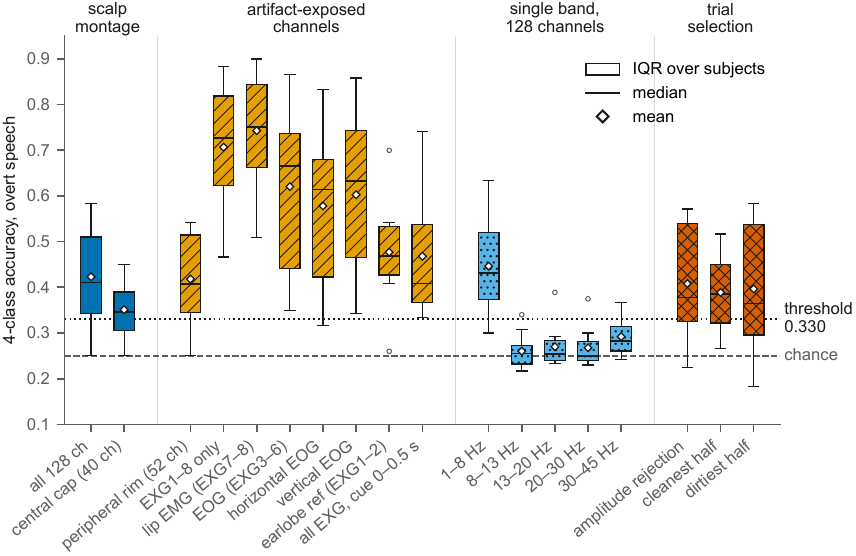}
\caption{Overt-speech decoding on ds003626 from restricted channel sets, bands and trial subsets. Four-class accuracy on pronounced trials, within subject, EEGNet ($F_1 = 16$, $D = 4$). Each box summarizes the ten participants: the box spans the interquartile range, the center line is the median, the diamond is the mean, whiskers extend to 1.5 times the interquartile range and circles mark participants beyond them. Fill and hatching identify the group of cells. The dashed line is chance and the dotted line the per-participant significance threshold. The threshold shown is that of the full trial set; the trial-selection cells test fewer trials and have thresholds of 0.333 to 0.364. The cell that applies the dataset authors' EMG criterion is omitted because it is available for only two participants.}
\label{fig:overt}
\end{figure}

\begin{figure}[t]
\centering
\includegraphics[width=\textwidth]{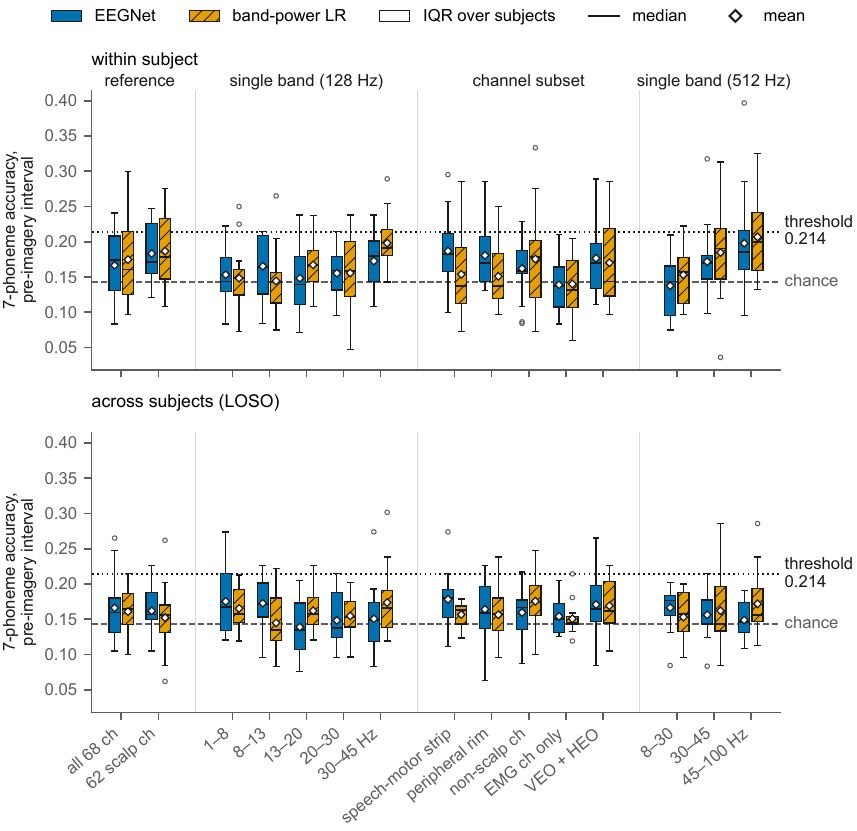}
\caption{Phoneme decoding in the KaraOne pre-imagery interval from channel subsets and single bands, for EEGNet (plain fill) and band-power logistic regression (hatched), within subject (top) and across participants (bottom). The rightmost group uses the 512\,Hz, 1--200\,Hz recordings. Each box summarizes the 14 participants, with the interval rebuilt as described in section~\ref{sec:karaone}, drawn as in figure~\ref{fig:overt}. The dashed line is chance and the dotted line the per-participant threshold.}
\label{fig:spoken}
\end{figure}

Table~\ref{tab:artifacts} summarizes the audit, and figures~\ref{fig:overt} and \ref{fig:spoken} show the full distributions over participants.

\begin{table}[t]
\caption{Artifact controls on the three positive controls, within subject. Values are mean accuracy over participants, measured in this study. For ds003626 the decoder is EEGNet ($F_1 = 16$, $D = 4$); for KaraOne and the competition corpus it is band-power logistic regression. The threshold is the exact binomial per-participant threshold at $\alpha = 0.05$. $^{\dagger}$Checked with its own shuffle-label control of 50 draws: the null mean lay between 0.2385 and 0.2409 against a chance level of 0.250, no draw reached the observed value, and $p = 1/51 \approx 0.020$ in every case.}
\label{tab:artifacts}
\centering
\begin{tabular}{@{}llcc@{}}
\toprule
Corpus and condition & Channels or band & Accuracy & Threshold \\
\midrule
ds003626, pronounced & 128-channel scalp montage$^{\dagger}$ & 0.4231 & 0.330 \\
 & central sensorimotor cap & 0.3509 & 0.330 \\
 & peripheral and temporal rim$^{\dagger}$ & 0.4181 & 0.330 \\
 & earlobe reference pair$^{\dagger}$ & 0.4769 & 0.330 \\
 & four EOG channels only$^{\dagger}$ & 0.6205 & 0.330 \\
 & eight EXG channels only$^{\dagger}$ & 0.7068 & 0.330 \\
 & two lip electrodes only$^{\dagger}$ & 0.7430 & 0.330 \\
 & 1--8\,Hz, all 128 channels & 0.4463 & 0.330 \\
 & 8--13, 13--20, 20--30\,Hz & 0.2601--0.2701 & 0.330 \\
 & 30--45\,Hz & 0.2919 & 0.330 \\[2pt]
KaraOne, pre-imagery, phonemes & 8--30\,Hz & 0.1531 & 0.214 \\
\quad (512\,Hz recordings) & 30--45\,Hz & 0.1847 & 0.214 \\
 & 45--100\,Hz & 0.2069 & 0.214 \\[2pt]
Competition, imagined phrases & 8--13\,Hz & 0.2638 & 0.235 \\
 & 20--30\,Hz & 0.3082 & 0.235 \\
 & 30--45\,Hz & 0.3560 & 0.235 \\
 & 62--100\,Hz & 0.3815 & 0.235 \\
\bottomrule
\end{tabular}
\end{table}

On ds003626, the 128-channel scalp montage decoded pronounced speech at 0.4231, with 8 of 10 participants above threshold. The eight EXG channels alone reached 0.7068 and the two lip electrodes alone 0.7430, with all ten participants above threshold in both cases, and the lip electrodes exceeded the full scalp montage by 0.32 (figure~\ref{fig:overt}). The peripheral rim outperformed the central sensorimotor cap (0.4181 against 0.3509). The mu and beta bands did not support decoding: each of the 8--13, 13--20 and 20--30\,Hz bands reached 0.2601 to 0.2701, with one participant of ten above threshold. The 1--8\,Hz band alone reached 0.4463, above the full band, and lies inside the frequency range that the dataset authors use to flag movement artifacts \cite{nieto2022}. The 30--45\,Hz band, where muscle activity dominates scalp recordings \cite{whitham2007}, reached 0.2919, above chance across participants ($p = 0.002$) although only two participants passed threshold. Even the earlobe pair, placed as a reference and not as a recording site, decoded at 0.4769. Trial selection by amplitude (0.4081), or by contrasting the cleanest with the dirtiest half of each participant's trials (0.3888 and 0.3965), left accuracy close to the full-montage value. The six channel sets marked in table~\ref{tab:artifacts}, from the scalp montage to the lip electrodes, each survived their own shuffle-label control; the central cap, the single bands and the trial-selection cells were not tested in this way.

In the KaraOne pre-imagery interval, participants move their articulators into position to pronounce the prompt \cite{zhao2015}, so a myogenic difference between prompts is expected there. Within subject, phoneme accuracy rose monotonically with frequency for both decoders: band power reached 0.1531, 0.1847 and 0.2069 in the 8--30, 30--45 and 45--100\,Hz bands, and EEGNet reached 0.1375, 0.1718 and 0.1979 (figure~\ref{fig:spoken}). Among the band-power cells, only the 30--45\,Hz band of the 128\,Hz recordings (0.1981) and the 45--100\,Hz band of the 512\,Hz recordings survived both FDR correction across the family and their own shuffle-label control, and no band within 8--30\,Hz survived correction for either decoder. Unlike ds003626, no single channel carried the effect, and the effect was not concentrated at the periphery: the EMG channel alone stayed below the scalp montage for both decoders under both protocols, and the speech-motor strip was not outperformed by the peripheral rim (0.1868 against 0.1807 for EEGNet within subject). Within subject, the evidence against a cortical origin on KaraOne is therefore the frequency profile alone.

Across participants the picture is mixed. With band power, the cells that survived FDR correction and their own shuffle-label control were the same two high-frequency bands, the six non-scalp channels and the ocular pair, all of them exposed to artifact, whereas the full scalp montage, the speech-motor strip, the 8--13\,Hz band and the 8--30\,Hz band of the 512\,Hz recordings failed their controls. With EEGNet, the 8--13\,Hz band (0.1726), the 8--30\,Hz band of the 512\,Hz recordings (0.1664) and the speech-motor strip (0.1781) survived FDR correction, but so did the ocular pair (0.1707), and no more than three of the 14 participants passed threshold in any of these cells. The 8--13\,Hz band, the speech-motor strip and the ocular pair each passed a 20-draw shuffle-label control: the null means lay between 0.1370 and 0.1432 against a chance level of 0.143, no draw reached the observed value, and $p = 1/21 \approx 0.048$ in each case, although the highest null draws (0.1647, 0.1737 and 0.1544) came closer to the observed values than in any ds003626 control. The 8--30\,Hz band of the 512\,Hz recordings was not tested in this way. With band power, the 1--8 and 13--20\,Hz bands also survived FDR correction across participants and passed their 200-draw controls ($p = 0.015$ and 0.035). A weak cross-participant effect in the mu and beta bands and over the speech-motor strip therefore survives correction and its controls, but so does an effect of similar size in the ocular channels. We therefore treat the KaraOne pre-imagery interval as evidence neither for nor against cortical speech-motor activity across participants.

On the competition corpus, which records imagined speech only, band-power accuracy followed the same progression, from 0.2638 in 8--13\,Hz to 0.3082 in 20--30\,Hz, 0.3560 in 30--45\,Hz and 0.3815 in 62--100\,Hz. This corpus has no EMG channel, so the inference rests on the frequency profile alone. Inner speech in ds003626 behaved differently: pairwise word accuracy fell from 0.5087 in the mu band to 0.4772 in 30--45\,Hz \cite{companionA2}, the opposite of the rise seen in the pronounced condition, the KaraOne pre-imagery interval and the competition's imagined phrases above. That profile is, however, measured on cells that are all at chance, and an absence of contamination is not evidence of content.

\subsection{Non-cortical channels survive channel selection}\label{sec:reference}

KaraOne stores 68 channels typed as EEG, six of which are not cortical: two mastoids, vertical and horizontal electro-oculography, electrocardiography and a dedicated EMG channel. Because the reference is the average of all 68, every stored channel equals its original value minus that average, and a subset chosen to exclude the non-cortical channels still carries each of them at $-1/68$ of its amplitude. Re-referencing the subset to its own mean removes this term exactly, and all KaraOne results here use the 62 scalp channels re-referenced in this way.

The effect of omitting this step is large. Decoding imagery versus rest across participants with band power over all 68 channels reached 0.9802, and the EMG channel alone accounted for 98.3\% of that above-chance accuracy. The same contrast on the re-referenced scalp channels is analyzed in the companion paper \cite{companionA2}. The phase features showed the same failure: without re-referencing, theta--beta coupling over the speech-motor region separated imagery from rest at 0.7045 within subject, above what band power achieves on the re-referenced data, and re-referencing reduced the same cell to 0.5027.

\subsection{Detection floors for amplitude effects}\label{sec:floor}

\begin{figure}[t]
\centering
\includegraphics[width=\textwidth]{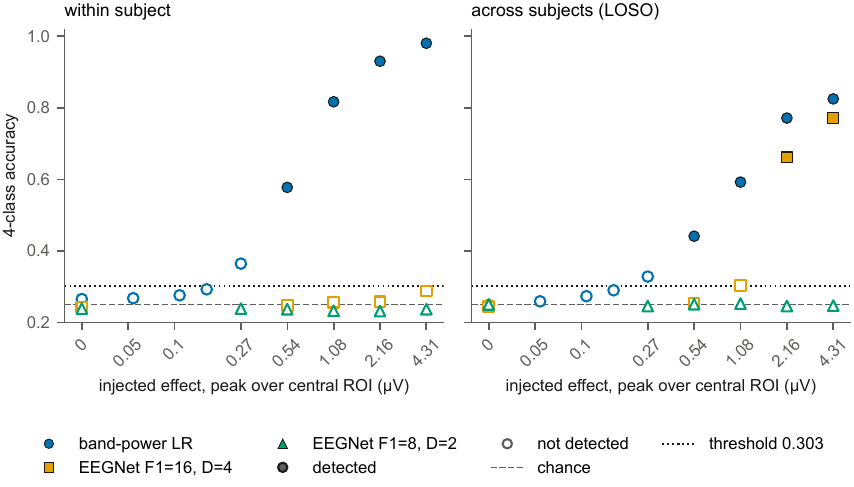}
\caption{Detection of an injected 8--30\,Hz desynchronization in real ds003626 inner-speech trials, within subject (left) and across participants (right). Points are mean accuracies; marker shape identifies the decoder, and a filled marker meets the pre-specified detection criterion. The effect amplitude is the peak reduction of the ongoing rhythm over the central sensorimotor region, on a logarithmic axis with the no-injection condition placed at 0. The dashed line is chance and the dotted line the per-participant threshold.}
\label{fig:floor}
\end{figure}

Table~\ref{tab:floor} and figure~\ref{fig:floor} give the accuracy of each decoder on the injected amplitude effect. Band-power logistic regression detected the effect from $k = 0.1$, a 0.54\uV\ desynchronization, under both protocols, and this held across three seeds. At $k = 0.05$ (0.27\uV) its mean accuracy exceeded the threshold but only seven of ten participants cleared it, so the pre-specified criterion was not met.

\begin{table}[t]
\caption{Mean four-class accuracy on ds003626 inner-speech trials with an injected 8--30\,Hz desynchronization of increasing amplitude. Bold values meet the pre-specified detection criterion (mean at or above the per-participant threshold of 0.303, at least eight of ten participants above it, and group Wilcoxon $p < 0.05$). Band-power values at $k$ = 0.05 to 0.2 are means over three seeds. A dash marks an amplitude that was not run for that configuration. Per-participant counts and $p$-values are given in supplementary table~S1.}
\label{tab:floor}
\centering
\begin{tabular}{@{}lcccccccc@{}}
\toprule
 & & \multicolumn{2}{c}{Band power} & \multicolumn{2}{c}{EEGNet, $F_1{=}16$, $D{=}4$} & \multicolumn{2}{c}{EEGNet, $F_1{=}8$, $D{=}2$} \\
\cmidrule(lr){3-4}\cmidrule(lr){5-6}\cmidrule(lr){7-8}
$k$ & Peak (\textmu V) & Within & LOSO & Within & LOSO & Within & LOSO \\
\midrule
0    & 0    & 0.2655 & 0.2488 & 0.2434 & 0.2439 & 0.2392 & 0.2505 \\
0.01 & 0.05 & 0.2679 & 0.2592 & -- & -- & -- & -- \\
0.02 & 0.11 & 0.2761 & 0.2736 & -- & -- & -- & -- \\
0.03 & 0.16 & 0.2927 & 0.2901 & -- & -- & -- & -- \\
0.05 & 0.27 & 0.3646 & 0.3284 & -- & -- & 0.2388 & 0.2462 \\
0.1  & 0.54 & \textbf{0.5777} & \textbf{0.4413} & 0.2476 & 0.2526 & 0.2372 & 0.2513 \\
0.2  & 1.08 & \textbf{0.8171} & \textbf{0.5925} & 0.2567 & 0.3037 & 0.2331 & 0.2530 \\
0.4  & 2.16 & \textbf{0.9304} & \textbf{0.7715} & 0.2579 & \textbf{0.6618} & 0.2322 & 0.2459 \\
0.8  & 4.31 & \textbf{0.9807} & \textbf{0.8251} & 0.2881 & \textbf{0.7718} & 0.2373 & 0.2471 \\
\bottomrule
\end{tabular}
\end{table}

EEGNet with $F_1 = 16$, $D = 4$ did not detect the effect within subject at any amplitude tested, reaching 0.2881 at 4.31\uV, while under LOSO it detected the effect from 2.16\uV. Its within-subject floor therefore lies above 4.31\uV, about eight times the band-power floor. With $F_1 = 8$, $D = 2$, EEGNet detected nothing under either protocol, including at 4.31\uV, where band power reached 0.9807 within subject. We re-ran the project's earlier EEGNet inner-speech analyses at the larger capacity with identical splits, and the conclusions held (within subject 0.2392 against 0.2434, LOSO 0.2505 against 0.2439), but only the larger configuration's results can be interpreted.

The difference is architectural and not only a matter of size. EEGNet applies an exponential linear unit \cite{clevert2016} followed by average pooling. The mean of a zero-mean oscillation is close to zero whatever its amplitude, and the exponential linear unit is nearly linear around zero, so an amplitude modulation is largely removed before the classifier sees it. ShallowConvNet squares, mean-pools and takes the logarithm, which is an explicit band-power estimate \cite{schirrmeister2017}, and on the same arrays, folds and training schedule it detected the effect at 2.16\uV\ under both protocols. The injected effect is an oscillatory amplitude change, the same class of effect that imagined speech is expected to produce in sensorimotor cortex \cite{pfurtscheller1999,jenson2014}, so this limitation applies to the intended target. We did not test EEGNet's sensitivity to phase-locked evoked responses.

The ordering between protocols is also informative. The injected effect is identical in both protocols by construction; what differs is the amount of training data, about 179 rows per within-subject fold against about 2,012 per LOSO fold. At $k = 0.4$ the three decoders order themselves by the number of parameters per training row (table~\ref{tab:capacity}). Band power, with 512 parameters, performed better within subject, whereas ShallowConvNet and EEGNet performed better across participants, with the largest reversal for the model with the most parameters per row. Accuracy under LOSO exceeding within-subject accuracy is usually treated as a sign of leakage; here the splits were asserted valid and the effect is the same in both arms, and the reversal reflects only the elevenfold difference in training data.

\begin{table}[t]
\caption{Accuracy on the same injected effect ($k = 0.4$, 2.16\uV) and the same folds, by decoder and protocol. Parameters per training row refer to the within-subject folds.}
\label{tab:capacity}
\centering
\setlength{\tabcolsep}{3pt}
\begin{tabular}{@{}lccccc@{}}
\toprule
Decoder & Parameters & Within subject & LOSO & Difference & Parameters per row \\
\midrule
Band-power logistic regression & 512 & 0.9304 & 0.7715 & $-$0.159 & 2.9 \\
ShallowConvNet & 6,885 & 0.4590 & 0.7370 & $+$0.278 & 39 \\
EEGNet ($F_1 = 16$, $D = 4$) & 17,188 & 0.2579 & 0.6618 & $+$0.404 & 96 \\
\bottomrule
\end{tabular}
\end{table}

\subsection{Detection floors across effect types and corpora}\label{sec:grid}

Table~\ref{tab:grid} extends the floor to four effect types and three corpora. Against each cell's own no-injection baseline, band power detected the amplitude effect from 0.27\uV\ on ds003626 and 0.24\uV\ on the competition corpus, and the Riemannian decoder from 0.54 and 0.49\uV. Floors on KaraOne were higher in microvolts, from 2.35\uV: the same $k$ corresponds to a larger amplitude there ($k = 0.1$ is 2.35\uV\ against 0.54\uV\ on ds003626), and KaraOne has fewer trials per class. Evoked deflections required several-fold larger amplitudes, from 3.39\uV\ on ds003626. Neither decoder detected class-dependent phase locking on any corpus at any injected size, and cross-frequency coupling was detected in only one of 60 cells. Amplitude-based and covariance-based decoders are therefore effectively blind to phase effects, which is why phase features were built and validated separately.

\begin{table}[t]
\caption{Smallest injected effect detected, by corpus, effect type and decoder, against the same cell without injection (paired Wilcoxon test, mean gain above one percentage point, Benjamini--Hochberg correction across all 240 cells). Amplitudes are peak values over the injected region in \textmu V; nd, not detected at any injected size (largest phase-locking and coupling injections are given in the text). The Riemannian decoder is tangent-space logistic regression after principal component reduction and re-centering.}
\label{tab:grid}
\centering
\begin{tabular}{@{}llcccc@{}}
\toprule
 & & \multicolumn{2}{c}{Band power} & \multicolumn{2}{c}{Riemannian} \\
\cmidrule(lr){3-4}\cmidrule(lr){5-6}
Corpus & Effect & Within & LOSO & Within & LOSO \\
\midrule
ds003626 & amplitude (8--30\,Hz) & 0.27 & 0.27 & 0.54 & 0.54 \\
 & evoked & 3.39 & 3.39 & 6.79 & 3.39 \\
 & phase locking & nd & nd & nd & nd \\
 & cross-frequency coupling & nd & nd & nd & $k = 0.8$ \\[2pt]
Competition & amplitude (8--30\,Hz) & 0.24 & 0.24 & 0.49 & 0.49 \\
 & evoked & 3.89 & 3.89 & 7.78 & 3.89 \\
 & phase locking & nd & nd & nd & nd \\
 & cross-frequency coupling & nd & nd & nd & nd \\[2pt]
KaraOne & amplitude (8--30\,Hz) & 2.35 & 4.70 & 9.40 & 4.70 \\
 & evoked & 5.00 & 5.00 & nd & 5.00 \\
 & phase locking & nd & nd & nd & nd \\
 & cross-frequency coupling & nd & nd & nd & nd \\
\bottomrule
\end{tabular}
\end{table}

The largest phase-locking injections raised inter-trial phase coherence from 0.118 to 0.391 on ds003626, from 0.099 to 0.255 on the competition corpus and from 0.254 to 0.363 on KaraOne, and the largest coupling injections raised the mean vector length from 0.091 to 0.176, 0.102 to 0.148 and 0.138 to 0.180. The two criteria used in this paper answer different questions. The pre-specified criterion of section~\ref{sec:floor} asks whether a decoder is above chance with most participants individually significant, which is the standard a reported positive result must meet; the grid criterion asks whether injection changes accuracy at all. The latter is more sensitive, and on ds003626 it places the band-power floor at 0.27\uV\ instead of 0.54\uV.

\subsection{Validation of the phase features}\label{sec:phase}

Before being applied to real content, the connectivity and coupling features were tested on effects built for them (table~\ref{tab:phasevalid}). On a pure zero-lag common source, the phase-locking value read 0.75 against 0.05 for the imaginary part of coherency. Under a common average reference with one high-amplitude channel, the phase-locking value rose by 0.48 while all three lagged measures moved slightly negative. Each validated family responded only to the effect built for it, and band power over the same region did not respond to either.

\begin{table}[t]
\caption{Largest analysis-of-variance $F$ statistic over classes for each feature family, over the same region, with and without injected effects that preserve band power.}
\label{tab:phasevalid}
\centering
\begin{tabular}{@{}lccc@{}}
\toprule
Injected effect & Connectivity (wPLI) & Theta--beta coupling & Band power \\
\midrule
none (real recordings) & 4.86 & 2.48 & 0.90 \\
inter-region phase lag & \textbf{1785.83} & 2.48 & 0.90 \\
theta--beta coupling & 4.85 & \textbf{12.08} & 0.89 \\
\bottomrule
\end{tabular}
\end{table}

Measured on all three corpora with the same folds and features used for real content, the detection floor was $k = 0.05$ to 0.10 for both validated families, with a single exception at $k = 0.40$ for coupling under LOSO on KaraOne (table~\ref{tab:phasefloor}). For the phase lag, $k$ is the fraction of one region's in-band component replaced by a lagged copy of the other's; for coupling it is the depth of modulation of the beta envelope by theta phase. Neither changes signal amplitude, so these floors have no microvolt equivalent. The third family, phase at a fixed latency, separated the classes with an $F$ statistic of 200 when the injected effect was not jittered and of 4 when it was. A per-trial latency jitter of 0.10\,s corresponds to 1.9 cycles at the center of the 8--30\,Hz band, so an effect locked to a jittered internal event is nearly unreadable at a fixed latency; detecting one would require a per-trial latency search, which we did not implement. We therefore treat that family as exploratory.

\begin{table}[t]
\caption{Detection floors for the two validated phase feature families. $k$ is the injected mechanism (lagged mixing fraction or modulation depth), not an accuracy. The last three columns are accuracy at the floor, accuracy with nothing injected, and the corpus's chance level.}
\label{tab:phasefloor}
\centering
\begin{tabular}{@{}lllcccc@{}}
\toprule
Corpus & Injected effect & Protocol & Floor ($k$) & At floor & At $k = 0$ & Chance \\
\midrule
ds003626 & inter-region phase lag & within & 0.1 & 0.3169 & 0.2483 & 0.2500 \\
 & & LOSO & 0.05 & 0.2921 & 0.2325 & 0.2500 \\
 & theta--beta coupling & within & 0.1 & 0.2852 & 0.2610 & 0.2500 \\
 & & LOSO & 0.05 & 0.2762 & 0.2568 & 0.2500 \\[2pt]
Competition & inter-region phase lag & within & 0.05 & 0.2410 & 0.2065 & 0.2000 \\
 & & LOSO & 0.05 & 0.2477 & 0.1918 & 0.2000 \\
 & theta--beta coupling & within & 0.1 & 0.2240 & 0.2135 & 0.2000 \\
 & & LOSO & 0.1 & 0.2218 & 0.2023 & 0.2000 \\[2pt]
KaraOne & inter-region phase lag & within & 0.05 & 0.2226 & 0.1400 & 0.1429 \\
 & & LOSO & 0.05 & 0.2291 & 0.1397 & 0.1429 \\
 & theta--beta coupling & within & 0.05 & 0.1412 & 0.1283 & 0.1429 \\
 & & LOSO & 0.4 & 0.2035 & 0.1347 & 0.1429 \\
\bottomrule
\end{tabular}
\end{table}

\section{Discussion}

The central finding is that the positive controls available in these corpora do not demonstrate that a pipeline decodes cortical speech activity. On ds003626 the evidence is direct: two electrodes over the lips outperform the full scalp montage, 0.7430 against 0.4231, nothing survives in the mu and beta bands, and even the earlobe reference decodes above threshold. On KaraOne and the competition corpus the evidence is spectral: within subject, accuracy climbs with frequency into the range where muscle activity dominates scalp recordings \cite{whitham2007,goncharova2003}. These are the signatures that reviews of muscle contamination describe \cite{muthukumaraswamy2013,mcmenamin2010}, and they appear in exactly the conditions used to argue that a pipeline works.

The failure is easy to miss for a structural reason. Overt speech is chosen as a positive control because it is expected to decode well, so a high accuracy is read as confirmation rather than as a reason to look further. Our own analyses treated the ds003626 overt result as a passing control for several months. For clinical translation the consequence is direct: a decoder validated on healthy speakers through articulator activity would fail for users who have no articulatory control, and the population for whom the technology is intended is the one in which this failure cannot be observed during development \cite{brumberg2010,vansteensel2023}.

The detection floors turn null results into bounded statements. On ds003626, a band-power decoder reliably detects a 10\% reduction of the ongoing mu and beta rhythm, 0.54\uV\ at its peak, and a more sensitive paired criterion lowers this to 0.27\uV. The floors also show that decoders differ sharply in what they can see. A compact convolutional network commonly used for EEG was blind within subject to an amplitude effect about eight times larger than the band-power floor, amplitude- and covariance-based decoders were blind to phase locking and cross-frequency coupling on every corpus, and phase features detected lagged mixing and coupling of 5--10\% but only because they were built for those effects. A null result therefore needs to name the decoder and the effect type it bounds. Intracranial studies suggest that imagined speech may be carried by low-frequency and cross-frequency features \cite{proix2022}, which are precisely the effects that band-power and Riemannian decoders did not detect here.

The ds003626 findings concern how a corpus is read, not errors in the deposited data. The condition mapping, the change of class encoding between releases and the blocked acquisition order are each documented in the dataset's own code and files, yet each can silently change a result. The collinearity of condition with time is a known hazard of block designs in EEG classification \cite{li2021} and deserves explicit mention wherever ds003626 is used to contrast conditions.

These observations lead to specific recommendations for scalp-EEG speech-decoding studies: report accuracy in the mu and beta bands alone; report accuracy from peripheral or artifact channels alone, and record EMG where the paradigm allows; report how accuracy changes with frequency, treating a monotonic rise into the range above 30\,Hz as evidence of contamination until shown otherwise; re-reference any channel subset to its own mean when the original reference included non-cortical channels; state a measured detection floor for each decoder and effect type used to support a null; and check label semantics and acquisition order against the source code of the dataset, and segment annotations against its published trial structure.

Several limitations apply. On KaraOne and the competition corpus the myogenic interpretation rests on the frequency profile: the competition corpus has no muscle channel, KaraOne's EMG channel alone does not carry the effect, and an attempt to locate it by recovering the decoders' per-channel weights failed our own consistency criterion (mean pairwise map correlation 0.078 to 0.118 against a required 0.30), so spatial evidence is missing, not contrary. A corpus recording masseter and laryngeal EMG alongside scalp EEG would allow a direct test. The ds003626 shuffle-label controls used 50 draws instead of 200, because each draw retrains the network for every participant, so their smallest attainable empirical $p$ is 0.020; no draw in any of the six controls came within 0.14 of the observed value. The three KaraOne EEGNet controls used 20 draws, for the same reason, so their smallest attainable $p$ is 0.048. The KaraOne audit concerns the pre-imagery interval, not speech, and KaraOne's spoken productions were not analyzed. Finally, each injected effect is a model of a real effect; the floors bound effects of the types injected, and a phase effect locked to a jittered internal event would escape the fixed-latency phase features.

\section{Conclusions}

On three public corpora used for imagined-speech decoding, the overt-speech and high-frequency effects that would ordinarily validate a scalp-EEG pipeline are dominated by muscle activity. Measured detection floors show that band power reliably detects a 0.54\uV\ mu and beta desynchronization on ds003626, that a widely used convolutional network does not detect it within subject at about eight times that amplitude, and that amplitude- and covariance-based decoders do not detect phase effects at all. Three features of ds003626, and a KaraOne annotation that labels an interval without speech as speaking, can silently alter results that use them. Reporting band-limited and artifact-channel accuracy, the frequency profile of accuracy and a measured detection floor would make positive and null results in this field considerably easier to interpret.

\ack{During the preparation of this work the authors used Claude (Anthropic) to assist with revising analysis and figure code, and with editing the manuscript text. The authors reviewed, validated and critically assessed all AI-assisted output, checked every reference against its source, and take full responsibility for the content of this article.
}

\funding{This research received no specific grant from any funding agency in the public, commercial or not-for-profit sectors.}

\roles{Blessing Isoyiza Adeika: conceptualization, methodology, software, formal analysis, investigation, data curation, visualization, writing -- original draft. Onyema Osuagwu: conceptualization, methodology, supervision, writing -- review and editing.}

\data{All three corpora are publicly available. ds003626 is available from OpenNeuro \cite{nieto2022data,markiewicz2021}, KaraOne from the University of Toronto \cite{zhao2015}, and the 2020 BCI Competition data from the competition organizers \cite{jeong2022}. Analysis code, per-run records, split hashes and the scripts that generate every table and figure are available at \url{https://github.com/blessadeik/scalp-eeg-speech-decoding}.
}

\suppdata{Supplementary material (PDF): the full detection-floor table with per-participant counts and group $p$-values for every decoder, protocol and injected amplitude (table~S1).}

\section*{Conflict of interest}
The authors declare no conflicts of interest.

\section*{Ethical statement}
This study analyzes only publicly released, de-identified datasets whose collection was approved by the ethics boards named in the original publications \cite{nieto2022,zhao2015,jeong2022}. No new data were collected, and no further approval was required for this secondary analysis.

\bibliographystyle{jne_iop_num}
\bibliography{references}

\clearpage
\section*{Supplementary material}
\setcounter{table}{0}
\renewcommand{\thetable}{S\arabic{table}}

\section*{S1. Full detection-floor table}
A class-dependent 8--30\,Hz desynchronization of known amplitude was injected into real ds003626 inner-speech trials. Detection requires the mean accuracy to reach the mean per-participant threshold (0.303), at least eight of ten participants to clear it individually, and a group Wilcoxon $p < 0.05$, a criterion fixed before the sweep; where several seeds exist, every seed must meet it, and the values shown are means over seeds.

\setlength{\tabcolsep}{3pt}
\begin{longtable}{@{}llccccccc@{}}
\caption{Accuracy, number of participants above threshold and group $p$ for every decoder, protocol and injected amplitude.}\label{tab:s1}\\
\toprule
Decoder & Protocol & $k$ & Peak (\textmu V) & Accuracy & $n_{\geq}$ & $p$ & Detected & Seeds \\
\midrule
\endhead
Band power & LOSO & 0 & 0 (none injected) & 0.2488 & 0/10 & 0.5898 & no & 1 \\
Band power & LOSO & 0.01 & 0.05 & 0.2592 & 1/10 & 0.2734 & no & 1 \\
Band power & LOSO & 0.02 & 0.11 & 0.2736 & 2/10 & 0.0371 & no & 1 \\
Band power & LOSO & 0.03 & 0.16 & 0.2901 & 3/10 & 0.0039 & no & 1 \\
Band power & LOSO & 0.05 & 0.27 & 0.3284 & 7/10 & 0.0013 & no & 3 \\
Band power & LOSO & 0.1 & 0.54 & 0.4413 & 9/10 & 0.0010 & yes & 3 \\
Band power & LOSO & 0.2 & 1.08 & 0.5925 & 10/10 & 0.0010 & yes & 3 \\
Band power & LOSO & 0.4 & 2.16 & 0.7715 & 10/10 & 0.0010 & yes & 1 \\
Band power & LOSO & 0.8 & 4.31 & 0.8251 & 10/10 & 0.0010 & yes & 1 \\
Band power & within & 0 & 0 (none injected) & 0.2655 & 1/10 & 0.1104 & no & 1 \\
Band power & within & 0.01 & 0.05 & 0.2679 & 1/10 & 0.0420 & no & 1 \\
Band power & within & 0.02 & 0.11 & 0.2761 & 1/10 & 0.0068 & no & 1 \\
Band power & within & 0.03 & 0.16 & 0.2927 & 3/10 & 0.0020 & no & 1 \\
Band power & within & 0.05 & 0.27 & 0.3646 & 7/10 & 0.0023 & no & 3 \\
Band power & within & 0.1 & 0.54 & 0.5777 & 9/10 & 0.0010 & yes & 3 \\
Band power & within & 0.2 & 1.08 & 0.8171 & 10/10 & 0.0010 & yes & 3 \\
Band power & within & 0.4 & 2.16 & 0.9304 & 10/10 & 0.0010 & yes & 1 \\
Band power & within & 0.8 & 4.31 & 0.9807 & 10/10 & 0.0010 & yes & 1 \\
EEGNet, $F_1{=}16$, $D{=}4$ & LOSO & 0 & 0 (none injected) & 0.2439 & 1/10 & 0.7217 & no & 1 \\
EEGNet, $F_1{=}16$, $D{=}4$ & LOSO & 0.1 & 0.54 & 0.2526 & 2/10 & 0.5000 & no & 1 \\
EEGNet, $F_1{=}16$, $D{=}4$ & LOSO & 0.2 & 1.08 & 0.3037 & 4/10 & 0.0654 & no & 1 \\
EEGNet, $F_1{=}16$, $D{=}4$ & LOSO & 0.4 & 2.16 & 0.6618 & 10/10 & 0.0010 & yes & 1 \\
EEGNet, $F_1{=}16$, $D{=}4$ & LOSO & 0.8 & 4.31 & 0.7718 & 10/10 & 0.0010 & yes & 1 \\
EEGNet, $F_1{=}16$, $D{=}4$ & within & 0 & 0 (none injected) & 0.2434 & 0/10 & 0.7266 & no & 1 \\
EEGNet, $F_1{=}16$, $D{=}4$ & within & 0.1 & 0.54 & 0.2476 & 0/10 & 0.6787 & no & 1 \\
EEGNet, $F_1{=}16$, $D{=}4$ & within & 0.2 & 1.08 & 0.2567 & 0/10 & 0.1641 & no & 1 \\
EEGNet, $F_1{=}16$, $D{=}4$ & within & 0.4 & 2.16 & 0.2579 & 1/10 & 0.2480 & no & 1 \\
EEGNet, $F_1{=}16$, $D{=}4$ & within & 0.8 & 4.31 & 0.2881 & 5/10 & 0.0312 & no & 1 \\
EEGNet, $F_1{=}8$, $D{=}2$ & LOSO & 0 & 0 (none injected) & 0.2505 & 0/10 & 0.3594 & no & 1 \\
EEGNet, $F_1{=}8$, $D{=}2$ & LOSO & 0.05 & 0.27 & 0.2462 & 0/10 & 0.7422 & no & 1 \\
EEGNet, $F_1{=}8$, $D{=}2$ & LOSO & 0.1 & 0.54 & 0.2513 & 0/10 & 0.3047 & no & 1 \\
EEGNet, $F_1{=}8$, $D{=}2$ & LOSO & 0.2 & 1.08 & 0.2530 & 0/10 & 0.2344 & no & 1 \\
EEGNet, $F_1{=}8$, $D{=}2$ & LOSO & 0.4 & 2.16 & 0.2459 & 0/10 & 0.6875 & no & 1 \\
EEGNet, $F_1{=}8$, $D{=}2$ & LOSO & 0.8 & 4.31 & 0.2471 & 0/10 & 0.7188 & no & 1 \\
EEGNet, $F_1{=}8$, $D{=}2$ & within & 0 & 0 (none injected) & 0.2392 & 0/10 & 0.8984 & no & 1 \\
EEGNet, $F_1{=}8$, $D{=}2$ & within & 0.05 & 0.27 & 0.2388 & 0/10 & 0.8984 & no & 1 \\
EEGNet, $F_1{=}8$, $D{=}2$ & within & 0.1 & 0.54 & 0.2372 & 0/10 & 0.9053 & no & 1 \\
EEGNet, $F_1{=}8$, $D{=}2$ & within & 0.2 & 1.08 & 0.2331 & 0/10 & 0.9629 & no & 1 \\
EEGNet, $F_1{=}8$, $D{=}2$ & within & 0.4 & 2.16 & 0.2322 & 0/10 & 0.9863 & no & 1 \\
EEGNet, $F_1{=}8$, $D{=}2$ & within & 0.8 & 4.31 & 0.2373 & 0/10 & 0.9980 & no & 1 \\
\bottomrule
\end{longtable}

\end{document}